\documentclass[conference]{IEEEtran}
\usepackage[top=1in, bottom=1in, left=0.78in, right=0.78in]{geometry}
\IEEEoverridecommandlockouts
\usepackage{cite}
\usepackage{amsmath,amssymb,amsfonts}
\usepackage{algorithmic}
\usepackage{graphicx}
\usepackage{textcomp}
\usepackage{xcolor}
\usepackage{booktabs}
\usepackage{multirow}
\usepackage{float}
\usepackage[skip=5pt]{caption}
\def\BibTeX{{\rm B\kern-.05em{\sc i\kern-.025em b}\kern-.08em
    T\kern-.1667em\lower.7ex\hbox{E}\kern-.125emX}}
\usepackage{titlesec}

\titlespacing*{\subsection}{0pt}{*1}{*0.5}

\begin{document}
\title{A Physiological-Model-Based Neural Network Framework for Blood Pressure Estimation from Photoplethysmography Signals}

\author{
    Yaowen Zhang\IEEEauthorrefmark{1}, 
    Libera Fresiello\IEEEauthorrefmark{2}, 
    Peter H. Veltink\IEEEauthorrefmark{1}, 
    % \IEEEmembership{Senior Member, IEEE}, 
    Dirk W. Donker\IEEEauthorrefmark{2}, 
    and Ying Wang\IEEEauthorrefmark{1}%
    \thanks{
        This work was supported in part by the scholarship from China Scholarship Council (CSC). 
        \IEEEauthorrefmark{1}Department of Biomedical Signals and Systems, University of Twente, The Netherlands. 
        Email: \{y.zhang-12, p.h.veltink, ying.wang\}@utwente.nl. 
        \IEEEauthorrefmark{2}Department of Cardiovascular and Respiratory Physiology, University of Twente, The Netherlands. 
        Email: \{l.fresiello, d.w.donker\}@utwente.nl.
    }
}
% \author{Yaowen Zhang, Libera Fresiello, Peter H. Veltink, \IEEEmembership{Senior Member, IEEE}, Dirk W. Donker, and Ying Wang}
% \thanks{This work was supported in part by the scholarship from China Scholarship Council (CSC).

% Yaowen Zhang, Peter H. Veltink and Ying Wang are with the Department of Biomedical Signals and Systems, University of Twente, The Netherlands (e-mail: y.zhang-12@utwente.nl, p.h.veltink@utwente.nl, ying.wang@utwente.nl).

% Libera Fresiello and Dirk W. Donker are with the Department of Cardiovascular and Respiratory Physiology, University of Twente, (e-mail:l.fresiello@utwente.nl, d.w.donker@utwente.nl)}
\maketitle

\begin{abstract}
Continuous blood pressure (BP) estimation via photoplethysmography (PPG) remains a significant challenge, particularly in providing comprehensive cardiovascular insights for hypertensive complications. This study presents a novel physiological model-based neural network (PMB-NN) framework for BP estimation from PPG signals, incorporating the identification of total peripheral resistance (TPR) and arterial compliance (AC) to enhance physiological interpretability. Preliminary experimental results, obtained from a single healthy participant under varying activity intensities, demonstrated promising accuracy, with a median standard deviation of 6.88 mmHg for systolic BP and 3.72 mmHg for diastolic BP. The median error for TPR and AC was 0.048 mmHg·s/ml and -0.521 ml/mmHg, respectively. Consistent with expectations, both estimated TPR and AC exhibited a reduction as activity intensity increased.
\end{abstract}

\section{Introduction}

Hypertension affects over 1.28 billion adults globally, yet only 21\% received adequate control \cite{who2023hypertension}. Effective and continuous blood pressure (BP) monitoring and management in daily life activities are crucial for high-risk populations with hypertension-related complications. Traditional cuff-based BP measurements were intermittent and unsuitable for continuous, real-time monitoring due to the discomfort caused by repeated cuff inflation and deflation. Recently, photoplethysmography (PPG) sensors had emerged as a non-invasive and continuous alternative tool for measuring microvascular blood volume changes, presenting promising potential for BP estimation in daily life scenarios \cite{Hu2022}.

Based on PPG signals, the pulse wave velocity (PWV)-based method gained prominence and popularity for BP estimation \cite{Hu2022}. Although PWV has proven to be a reliable method for BP estimation, it requires two sensors placed at different locations, limiting its applicability in mobile health settings. To address this limitation, single PPG-sensor-based methods emerged, reducing hardware complexity and enhancing usability. These PPG-based methods include parametric models, which applies predefined physiological modeling equations and specific features (e.g., heart rate, systolic/diastolic intervals)\cite{Wang2018}, and non-parametric models, which relies solely on features derived from PPG signals \cite{Wang2018}. While parametric models provide insights into physiology, they often require high-quality PPG signals under well controlled environment and individual calibration for personalized estimations. This was impractical in dynamic daily life conditions \cite{Wang2018}\cite{El-Hajj2020}. Non-parametric models, such as neural networks \cite{Casadei2022}\cite{Sideris2016} and regression models \cite{Teng}\cite{Duan2016}\cite{Dey2018}, provide higher accuracy in real-life environments, though they are still affected by signal degradation from motion and lighting variations. The parametric models provide physiological insights, while the non-parametric models offer high accuracy, making them complementary for BP estimation. Accordingly, we hypothesized that the synergy between these two types of models can enable reliable and accurate BP estimation across diverse real-world conditions.

Hypertension is a key risk factor for coronary artery disease (CAD) due to its role in accelerating atherosclerosis. Increased total peripheral resistance (TPR) and reduced arterial compliance (AC), both associated with vessel stiffening, elevate cardiovascular stress and pulse pressure, particularly in elderly hypertensive patients \cite{Steppan2011}. Estimating TPR and AC concurrently with BP during daily life could have enhanced cardiovascular risk assessment, offered insights into therapeutic efficacy and paved the way for personalized hypertension management. While invasive catheterization had been the gold standard for assessing TPR and AC, non-invasive options, such as oscillometry combined with impedance cardiography \cite{Olano2019}, provide a feasible clinical alternative, though they are impractical for daily monitoring. In this study, we accordingly proposed a novel physiological model-based neural network (PMB-NN) to estimate BP from finger PPG while simultaneously identify TPR and AC across different activity intensities. 

\section{Methods}
\begin{figure*}
\centerline{\includegraphics[width=0.95\linewidth]{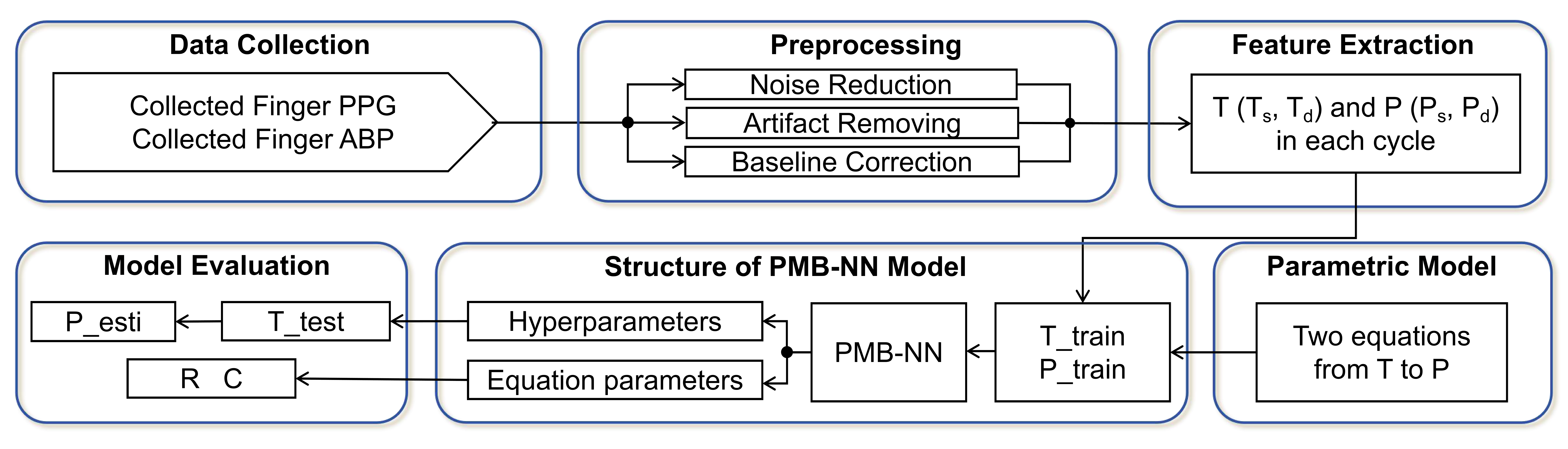}}
\captionsetup{skip=5pt}
\caption{Flowchart of validation procedure for PMB-NN.}
\label{fig1}
\end{figure*}
\subsection{Data collection}
A flow diagram illustrating the procedure, beginning with data collection for model training and testing, is shown in Figure \ref{fig1}. The study was approved by the Ethics Committee Computer and Information Science (EC-CIS) of the University of Twente (No. 240831) and conducted in Roessingh Research and Development (RRD). 10 healthy participants aged between 20 and 29 were enrolled in the study. Informed consent was received from all the participants and the study was performed in accordance with the declaration of Helsinki.

Participants were asked to perform three exercise activities in 35 minutes. They started with sitting 5 minutes on a chair and standing 5 minutes. Then they were asked to cycle on an Bremshey® exercise bike at a fixed speed of 45 rpm, power of 50($\pm$10) watts for 10 minutes. After 1 minute's resting, they started cycling again on the same bike at a fixed speed of 45 rpm, power of 100($\pm$10) watts for 10 minutes. Exercise intensities were set to assure patients trained within the aerobic threshold, and 10 minutes period length was chosen to guarantee the blood pressure, total peripheral resistance, and arterial compliance reached a steady-state, free from transient fluctuations at the start of exercise. PPG signal was collected on the left index finger (Shimmer 3 GSR+ Unit, Shimmer Wearable Sensory Technology) and BP was measured continuously using volume-clamp PPG on the left middle finger (Finometer® Model-2, Finapres Medical Systems, Enschede, the Netherlands). During the whole study, the left hand was placed in a sling suspended around the neck where the left hand’s index and middle fingers were kept at heart level to prevent hydrostatic pressure artifacts for blood pressure and PPG’s reliable measurement. Each participant has conducted the whole experiment twice on different days for training and testing sets obtaining.

\subsection{Data Preprocessing}
Beat-to-beat systolic and diastolic blood pressure (Ps, Pd) signals,  cardiac output (CO in mL/min) and total peripheral resistance (R in mmHg·s/ml) were exported from BeatScope® Easy software with Modelflow algorithm embeded on the beat-to-beat arterial waveform analysis. The pressure decay method \cite{liu1986estimation} was applied to each cycle's ABP waveform to extract arterial compliance (C in ml/mmHg). A second order Savitzky-Golay filter with the window size of 31 cycles was applied to smoothen the irregular fluctuation of the above variables (Ps, Pd, CO, R, C).

Finger PPG signals from the Shimmer GSR+ unit were exported via Consensys® software at 64 Hz. Pre-processing involved four steps: (1) linear detrending to remove baseline drift; (2) a 0.5-s median filter to suppress motion artifacts; (3) a fourth-order Butterworth bandpass filter (0.5–5.0 Hz) to isolate physiological components, with cutoffs selected based on the Nyquist frequency; and (4) a 0.1-s moving average filter for final smoothing.

\subsection{PMB-NN Inputs and Outputs}
\begin{figure}[H]
\centerline{\includegraphics[width=0.95\linewidth]{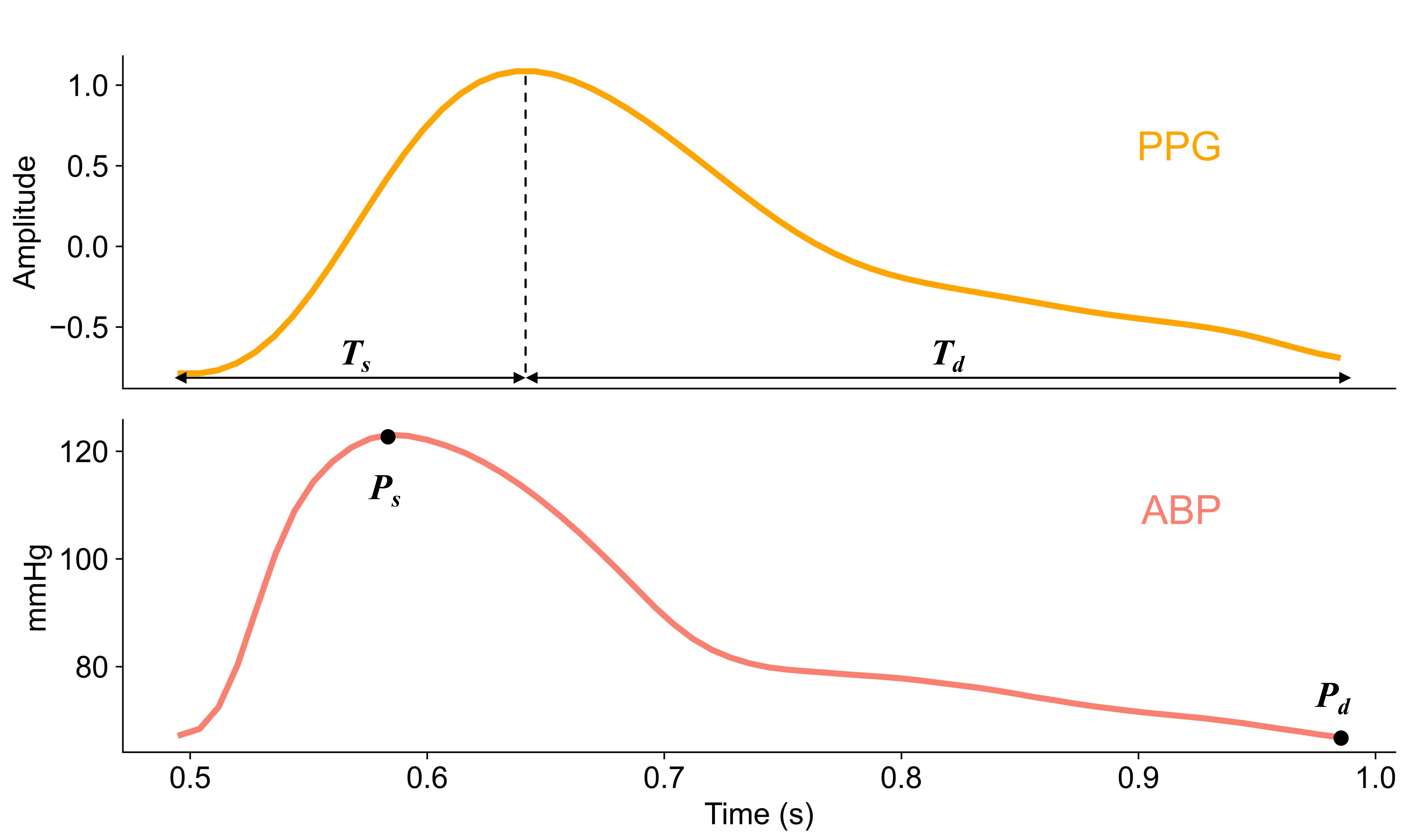}}
\captionsetup{skip=5pt}
\caption{$T_s$, $T_d$, $P_s$ and $P_d$ extraction from one cycle's PPG and ABP signal.}
\label{fig2}
\end{figure}

The systolic upstroke time ($T_s$) and diastolic time ($T_d$) from PPG were set as model inputs, while systolic pressure ($P_s$) and diastolic pressure ($P_d$) were set as outputs. Figure \ref{fig2} illustrated these variables in one cycle’s PPG and BP signal. For signal synchronization, both BP and PPG preprocessed signals were aligned according to the time stamps of each activity and each device’s system time. Since PPG and BP were collected at the middle knuckle of index and middle finger on left hand, respectively, the physiological blood-flow time difference measured by the two signals was assumed as negligible. The PMB-NN input vector $T$ is merged of the two feature arrays ($T_s$ and $T_d$) alternatively and the output vector $P$ is merged in the same way, either.

\subsection{Physiological Model}
The physiological model was defined by a set of validated equations (Eq.\ref{eq1} and Eq.\ref{eq2}) \cite{Sinha2012} based on the two-element Windkessel model proposed by Otto Frank \cite{Sagawa1990}, which describes the mechanical properties of the arterial vessels in terms of total peripheral resistance and arterial compliance. 
\begin{equation}
\begin{split}
P_{s,i} &= P(t | t = T_{s,i}) \\
    &= P_{d,i-1} e^{-T_{s,i} / RC} 
    + \frac{I_{0,i} T_{s,i} C \pi R^2}{T_{s,i}^2 + C^2 \pi^2 R^2} \left( 1 + e^{-T_{s,i} / RC} \right)
\end{split}
\label{eq1}
\end{equation}
\begin{equation}
P_{d,i} = P(t | t = T_{d,i}) = P_{s,i} e^{-T_{d,i} / RC}
\label{eq2}
\end{equation}
where the units for $T_{s,i}$ and $T_{d,i}$ are in seconds and for $P_{s,i}$ and $P_{d,i}$ are in mmHg. Every cardiac cycle is assumed to start with a systolic period, and $P_{d,0}$ is set to an initial value as the condition of first cardiac cycle in the current measurement. $R$ and $C$, in all the following text, refer to TPR and AC in units of mmHg·s/ml and ml/mmHg, respectively. The blood flows from the 
ventricle to the aorta was assumed as a sine wave where $I_0$ is the peak amplitude, calculated as Eq.\ref{eq3}.
\begin{equation}
\
I_{0,i} = \frac{CO_i \times T_{c,i}}{\int_0^{T_{s,i}} \sin\left(\frac{\pi t}{T_{d,i}}\right) dt}
\
\label{eq3}
\end{equation}
where $CO_i$ is cardiac output (in ml/s) and $T_{c,i}$ (= $T_{s,i}$ + $T_{d,i}$) is the duration of the $ith$ heart cycle.

Eq.\ref{eq1} and \ref{eq2}, as our physiological model, representing a continuous simulated correlation from $T_s$/$T_d$ to $P_s$/$P_d$. It is applied to add physiological constraint in neural network structure.

\subsection{Physiological Model-based Neural Network Structure}
\begin{figure*}[t]
\centerline{\includegraphics[width=0.95\linewidth]{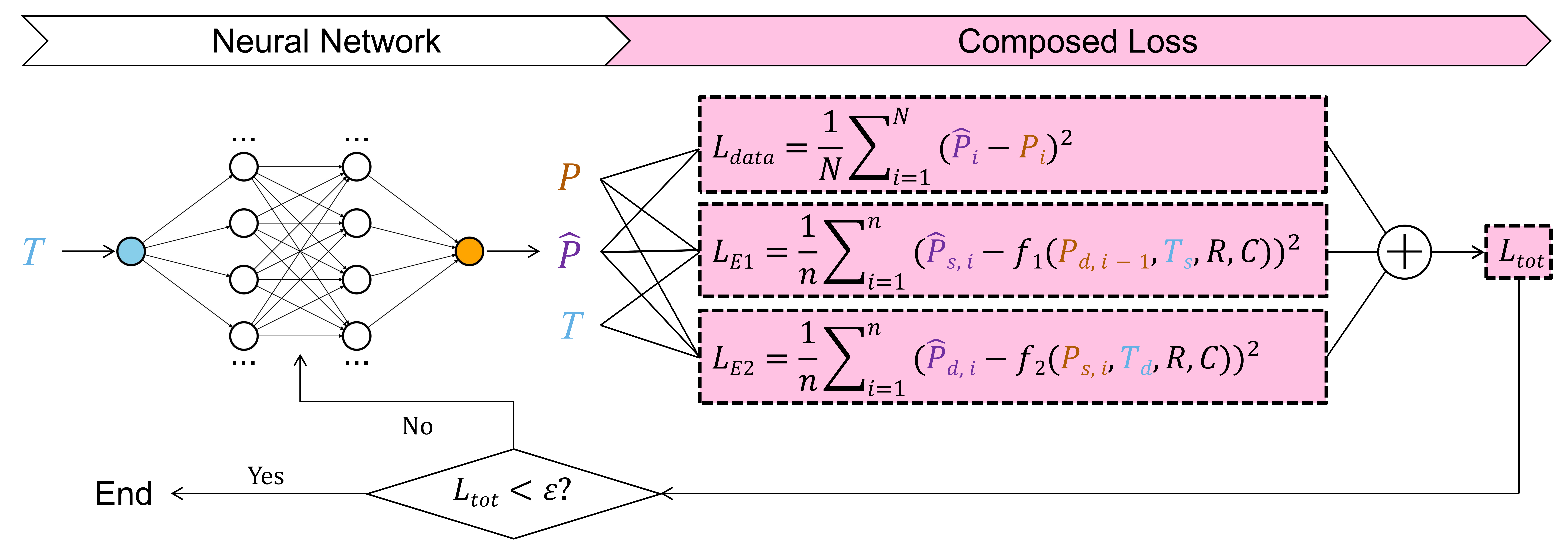}}
\captionsetup{skip=5pt}
\caption{PMB-NN structure diagram}
\label{fig3}
\end{figure*}

Figure \ref{fig3} reveals the structure of our PMB-NN model. It consists of two parts, neural network and composed loss. We constructed a fully connected neural network (FCNN) with single input T and single output P, and 1 input layer, 3 hidden layers, and1 output layer (1-128-128-128-1) with ReLU activation functions mapping the input $T$ (which comprises $T_s$ and $T_d$) to the output $\hat{P}$.  

The composed loss, consists of one data fitting term and two physiological constraint terms derived from Eq.\ref{eq1} and Eq.\ref{eq2}, was used to optimize the hyperparameters (weights and biases) in the FCNN training. The data fitting term, $L_{data}$ is calculated as a $L_2$ loss function in Eq.\ref{eq4}.
\begin{equation}
L_{data}=\frac{1}{N}\sum_{i=1}^{N} (\hat{P_i}-P_i)^2
\label{eq4}
\end{equation}
where $N$ is the length of $P$. $\hat{P}_i$ and $P_i$ are the FCNN output and ground truth of blood pressure, respectively. The physiological constraint terms, $L_{E1}$ and $L_{E2}$, are defined as another two $L_2$ loss functions in Eq.\ref{eq5} and \ref{eq6} according to the parametric model in Eq.\ref{eq1} and \ref{eq2}. Initial values for $R$ and $C$ were both set as 1.
\begin{equation}
L_{E1}=\frac{1}{n}\sum_{i=1}^{n} (\hat{P}_{s,i}-f_1(P_{d,i-1},T_s,R,C))^2
\label{eq5}
\end{equation}
\begin{equation}
L_{E2}=\frac{1}{n}\sum_{i=1}^{n} (\hat{P}_{d,i}-f_2(P_{s,i},T_d,R,C))^2
\label{eq6}
\end{equation}
where $n$ equals to $0.5N$, is the length of $P_s$/$P_d$. $f_1$ and $f_2$ are the functions at the right side of Eq.\ref{eq1} and \ref{eq2}. $\hat{P}_{s,i}$ and $\hat{P}_{d,i}$ represent the predicted systolic and diastolic pressure in the $i^{th}$ cycle. $P_{d,i-1}$ and $P_{s,i}$ are real diastolic pressure in the ${i-1}^{th}$ cycle and systolic pressure in the $i^{th}$ cycle. 

The optimization adjusts the model's parameters, including neural network weights, biases, and physiological parameters $R$ and $C$, to minimize the total loss, $L_{tot}$. The Adam optimizer, combined with a learning rate scheduler with initial rate as 0.01, was used to iteratively refine these parameters and ensure stable convergence. The model was iteratively trained until $L_{tot}$ drops below a predefined threshold of 10 mmHg² within 1000 iterations, ensuring the model achieves a balance between accuracy and physical plausibility.

\subsection{Model Evaluation}
Training and testing data sets were the same length (10 minutes each activity for different types) collected on the same participant on different dates, respectively. We tested the performance of PMB-NN by comparing estimated and measured pressure values counted on systolic pressure and diastolic pressure, separately. Both performances were revealed in terms of mean error (ME) (in mmHg) and standard deviation (SD) (in mmHg). The estimated $R$ was compared with the golden standard $R$ which was calculated by the Beatscope software, while we only reported the estimated $C$ given the lack of measured $C$ using Finometer Model-2 devices.

\section{Results}
We presented the preliminary result of only one participant's data. Table \ref{tab1} listed the error of PMB-NN estimation on systolic pressure and diastolic pressure, respectively. The estimation error for both $P_s$ and $P_d$ is the lowest in activity of resting among the three activities.

\begin{table}[H]
\centering
\caption{ME and SD values for Estimated Pressure in Different Activities}
\begin{tabular}{c|c|c|c|c}
\toprule
\multicolumn{1}{c|}{} & \multicolumn{4}{c}{\text{Estimated Pressure}}\\
\cmidrule{2-5}
\text{Activity}& \multicolumn{2}{c|}{\text{$P_s$}} & \multicolumn{2}{c}{\text{$P_d$}}\\
\cmidrule{2-5}
                          & \text{ME} & \text{SD} & \text{ME} & \text{SD}\\
\midrule
Resting                   & -2.81      & 6.62     & -4.12      & 4.35\\
Cycling 50 watts          & -4.18       & 7.37      & 3.1        & 3.72\\
Cycling 100 watts         & 5.7       & 6.88      & -0.32        & 2.66\\
\bottomrule
\end{tabular}
\label{tab1}
\end{table}

\begin{figure}
\centerline{\includegraphics[width=0.95\linewidth]{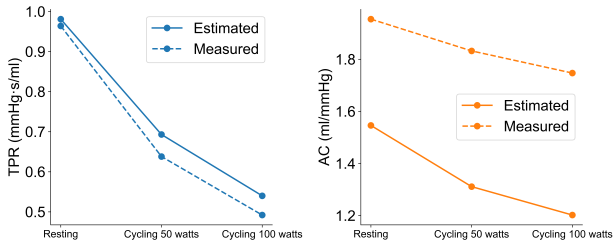}}
\captionsetup{skip=5pt}
\caption{Relevance plots of estimated and true values for R and C under three activity senarios. Solid lines represent estimated values; dashed lines represent measured values.}
\label{fig4}
\end{figure}

The measured and estimated $R$, $C$ from PMB-NN corresponding to various activity types are listed in figure \ref{fig4}. It showed that the difference between the estimated and measured $R$ is relatively small (median (min, max) as 0.048 (0.017, 0.055) mmHg·s/ml), while the error of $C$ estimation is larger (median (min, max) as -0.521 (-0.408, -0.545) ml/mmHg). We also observed that both the measured and estimated $R$ and $C$ decreased from resting to increasing cycling workload.

\section{Discussion}
Experimental results demonstrated that the PMB-NN model achieved accurate Ps estimation during both resting and 50-watt cycling, as well as Pd estimation across varying activity intensities. Both estimations met the Association for the Advancement of Medical Instrumentation (AAMI) standard, with a ME within $\pm$5 mmHg and a SD below 8 mmHg \cite{obrien2001blood}. Compared to the benchmark CNN LSTM method \cite{mahardika2023ppg}, which yielded a median SD of 6.88 (6.63, 7.19) mmHg for Ps estimation and 2.92 (1.99, 3.2) mmHg for P estimation on our dataset, the PMB-NN model demonstrated comparable performance. Furthermore, the estimated vascular resistance and compliance exhibited physiologically consistent trends, with R and C decreasing as activity intensity increased, aligning with established cardiovascular responses to exercise. These findings validate the model’s capability to capture dynamic hemodynamic adaptations.

The above findings proved that the PMB-NN framework holds potential for mobile health applications, particularly in high-risk populations requiring continuous BP monitoring. Its capability to estimate R and C alongside BP provides additional metrics for cardiovascular risk assessment, enabling a more comprehensive understanding of patient health. Furthermore, this approach could facilitate the early detection of hypertension-related complications and improve the personalization of treatment strategies by providing accurate BP, R, and C estimates.

Despite the promising results, it was notable that systolic pressure estimation was affected by higher errors compared to diastolic pressure. This discrepancy may stem from the inherent complexity and variability of systolic pressure due to transient cardiovascular dynamics during activity. Meanwhile, the estimated systolic and diastolic pressure values tended to converge toward the median, likely due to limited input variability, which failed to capture subtle blood pressure fluctuations. On the other hand, if the initial values for resistance and compliance are set orders of magnitude away from the true values, the model may struggle to converge quickly. Additionally, the unverified identification of the compliance parameter value (C) further limited the model's reliability. To address these limitations, future efforts should focus on incorporating additional signal features, such as waveform morphology, which may enhance the model's sensitivity to systolic fluctuations. To transition the PMB-NN framework from an experimental setting to a practical, real-world application, several key steps must be undertaken. We also plan to expand the dataset by incorporating data from 10 participants, which will further validate the PMB-NN framework and provide a more diverse range of data for improving model accuracy. This expanded dataset will allow for a deeper analysis of the model's performance across various demographic groups and health conditions, thereby enhancing its robustness. Additionally, this effort will pave the way for its potential application in mobile health monitoring systems, where continuous, non-invasive monitoring of BP, R, and C can provide valuable real-time insights into cardiovascular health, helping to personalize treatment strategies and enable early detection of complications.

\section{Conclusion}
This study introduced a novel PMB-NN framework for blood pressure estimation using PPG signals. By combining physiological model and deep learning approaches, the PMB-NN achieved promising performance under human movement conditions while providing physiological insights through TPR and AC estimation. The PMB-NN framework represented its potential toward daily BP monitoring. Its ability to deliver accurate BP estimates and additional cardiovascular metrics highlights its utility in mobile health applications like personalized hypertension management. 
\bibliographystyle{IEEEtran}
\bibliography{main}

\begin{thebibliography}{10}
\providecommand{\url}[1]{#1}
\csname url@samestyle\endcsname
\providecommand{\newblock}{\relax}
\providecommand{\bibinfo}[2]{#2}
\providecommand{\BIBentrySTDinterwordspacing}{\spaceskip=0pt\relax}
\providecommand{\BIBentryALTinterwordstretchfactor}{4}
\providecommand{\BIBentryALTinterwordspacing}{\spaceskip=\fontdimen2\font plus
\BIBentryALTinterwordstretchfactor\fontdimen3\font minus \fontdimen4\font\relax}
\providecommand{\BIBforeignlanguage}[2]{{%
\expandafter\ifx\csname l@#1\endcsname\relax
\typeout{** WARNING: IEEEtran.bst: No hyphenation pattern has been}%
\typeout{** loaded for the language `#1'. Using the pattern for}%
\typeout{** the default language instead.}%
\else
\language=\csname l@#1\endcsname
\fi
#2}}
\providecommand{\BIBdecl}{\relax}
\BIBdecl

\bibitem{who2023hypertension}
\BIBentryALTinterwordspacing
{World Health Organization}, ``Hypertension,'' 2023, accessed: 2025-04-21. [Online]. Available: \url{https://www.who.int/news-room/fact-sheets/detail/hypertension}
\BIBentrySTDinterwordspacing

\bibitem{Hu2022}
Q.~Hu, D.~Wang, and C.~Yang, ``Ppg-based blood pressure estimation can benefit from scalable multi-scale fusion neural networks and multi-task learning,'' \emph{Biomedical Signal Processing and Control}, vol.~78, p. 103891, 9 2022.

\bibitem{Wang2018}
G.~Wang, M.~Atef, and Y.~Lian, ``Towards a continuous non-invasive cuffless blood pressure monitoring system using ppg: Systems and circuits review,'' \emph{IEEE Circuits and Systems Magazine}, vol.~18, pp. 6--26, 2018.

\bibitem{El-Hajj2020}
C.~El-Hajj and P.~A. Kyriacou, ``A review of machine learning techniques in photoplethysmography for the non-invasive cuff-less measurement of blood pressure,'' \emph{Biomedical Signal Processing and Control}, vol.~58, p. 101870, 4 2020.

\bibitem{Casadei2022}
B.~C. Casadei, A.~Gumiero, G.~Tantillo, L.~D. Torre, and G.~Olmo, ``Systolic blood pressure estimation from ppg signal using ann,'' \emph{Electronics}, vol.~11, p. 2909, 9 2022.

\bibitem{Sideris2016}
C.~Sideris, H.~Kalantarian, E.~Nemati, and M.~Sarrafzadeh, ``Building continuous arterial blood pressure prediction models using recurrent networks.''\hskip 1em plus 0.5em minus 0.4em\relax IEEE, 5 2016, pp. 1--5.

\bibitem{Teng}
X.~Teng and Y.~Zhang, ``Continuous and noninvasive estimation of arterial blood pressure using a photoplethysmographic approach.''\hskip 1em plus 0.5em minus 0.4em\relax IEEE, pp. 3153--3156.

\bibitem{Duan2016}
K.~Duan, Z.~Qian, M.~Atef, and G.~Wang, ``A feature exploration methodology for learning based cuffless blood pressure measurement using photoplethysmography.''\hskip 1em plus 0.5em minus 0.4em\relax IEEE, 8 2016, pp. 6385--6388.

\bibitem{Dey2018}
J.~Dey, A.~Gaurav, and V.~N. Tiwari, ``Instabp: Cuff-less blood pressure monitoring on smartphone using single ppg sensor.''\hskip 1em plus 0.5em minus 0.4em\relax IEEE, 7 2018, pp. 5002--5005.

\bibitem{Steppan2011}
J.~Steppan, V.~Barodka, D.~E. Berkowitz, and D.~Nyhan, ``Vascular stiffness and increased pulse pressure in the aging cardiovascular system.'' \emph{Cardiology research and practice}, vol. 2011, p. 263585, 2011.

\bibitem{Olano2019}
R.~D. Olano, W.~G. Espeche, M.~R. Salazar, P.~Forcada, J.~A. Chirinos, A.~de~Iraola, C.~E.~L. Sisnieguez, B.~C.~L. Sisnieguez, E.~Balbín, and H.~A. Carbajal, ``Evaluation of ventricular-arterial coupling by impedance cardiography in healthy volunteers,'' \emph{Physiological Measurement}, vol.~40, p. 115002, 12 2019.

\bibitem{liu1986estimation}
\BIBentryALTinterwordspacing
Z.~Liu, K.~P. Brin, and F.~C. Yin, ``Estimation of total arterial compliance: an improved method and evaluation of current methods,'' \emph{The American Journal of Physiology}, vol. 251, no. 3 Pt 2, pp. H588--H600, 1986. [Online]. Available: \url{https://doi.org/10.1152/ajpheart.1986.251.3.H588}
\BIBentrySTDinterwordspacing

\bibitem{Sinha2012}
V.~V. M.~C. Mridu~Sinha, ``Model of aortic blood flow using the windkessel effect,'' 10 2012.

\bibitem{Sagawa1990}
K.~Sagawa, ``Translation of otto frank's paper “die grundform des arteriellen pulses” zeitschrift für biologie 37: 483–526 (1899),'' \emph{Journal of Molecular and Cellular Cardiology}, vol.~22, pp. 253--254, 3 1990.

\bibitem{obrien2001blood}
\BIBentryALTinterwordspacing
E.~O'Brien, B.~Waeber, G.~Parati, J.~Staessen, and M.~G. Myers, ``Blood pressure measuring devices: recommendations of the european society of hypertension,'' \emph{BMJ (Clinical research ed.)}, vol. 322, no. 7285, pp. 531--536, 2001. [Online]. Available: \url{https://doi.org/10.1136/bmj.322.7285.531}
\BIBentrySTDinterwordspacing

\bibitem{mahardika2023ppg}
\BIBentryALTinterwordspacing
T.~N.~Q. Mahardika, Y.~N. Fuadah, D.~U. Jeong, and K.~M. Lim, ``Ppg signals-based blood-pressure estimation using grid search in hyperparameter optimization of cnn-lstm,'' \emph{Diagnostics (Basel, Switzerland)}, vol.~13, no.~15, p. 2566, 2023. [Online]. Available: \url{https://doi.org/10.3390/diagnostics13152566}
\BIBentrySTDinterwordspacing

\end{thebibliography}

\end{document}